\documentclass{elsart}
\usepackage{amsfonts}
\usepackage{amssymb}
\usepackage{amsmath}

\input{tcilatex}
\begin{document}

\begin{center}
{\Large Econophysics: A new discipline}

\bigskip 

\textbf{S\'{o}nia R. Bentes}$^{\ast }$

ISCAL, Av. Miguel Bombarda 20, 1069-035 Lisboa, Portugal

20 June 2010

\bigskip 

\bigskip 

{\large \textbf{Abstract}}
\end{center}

This paper debates the contribution of Econophysics to the economic or
financial domains. Since the traditional approach performed by Economics or
Finance has revealed to be insufficient in fully characterizing and
explaining the correspondingly phenomena, we discuss whether Econophysics
can provide a new insight onto these matters. Thus, an assessment is
presented in order to weight its potential opportunities and limitations.
This is particularly relevant as it is widely recognized that during its yet
short existence Econophysics has experienced a growing interest not only by
physicists but also by economists in searching for new approaches that could
help explaining existing questions. In fact, many papers have been
submitted, some books have been released, new journals have been published,
several conferences have been held, a site is maintained --
http://www.unifr.ch/econophysics where news, events, book reviews, papers
and a blog are exhibited; a 3-year licentiate studies (University of Silesia
[1]) and a B.Sc. course (University of Wroclaw [2]) have been created and
also some Ph.D. thesis have been written. Therefore, a fundamental question
arises: Is this just a fad or is it something much more consistent that will
prevail? This is what this paper addresses.

\bigskip



$^{\ast }$E-mail: smbentes@iscal.ipl.pt

\section{Origins}

In the last decade physicists have been increasingly concerned about
economic or financial subjects. This has given way to a new branch of
knowledge called \textquotedblleft Econophysics\textquotedblright . This
neologism was first introduced by H.E. Stanley, in 1996 [3], in an attempt
to legitimize the study of Economics by physicists and is a result of the
combination of the word \textquotedblleft Economics\textquotedblright\ with
the word \textquotedblleft Physics\textquotedblright , in a clear analogy
with terms such as Biophysics, Geophysics and Astrophysics. Nevertheless,
the involvement of Physics in Social Sciences is longer, dating back at
least to 1526 when Copernicus, while studying the behaviour of the
inflation, established the theoretical foundations of what was later known
as the Greshnam law, according to what \textquotedblleft bad money drives
out good under legal tender laws\textquotedblright . Newton was the
following physicist involved in Economics, who as a warden of the Royal Mint
of Great Britain during 1969-1701, standardized Britain's coinage. Another
interesting insight was brought up by Edmond Halley, the famous
mathematician and astronomer, who has derived the foundations of life
insurance. Based on data relating to births and deaths in the German city of
Breslau, between 1687 and 1691, Halley constructed in 1693 his own life
table (for individual ages, not for groups), which was found to give a
reasonably accurate picture of survival and become well known throughout
Europe. Later, in 1738, Bernoulli introduced the idea of utility to describe
people's preferences. Subsequently, Laplace stressed out, in 1812, that
events that might seem random and unpredictable can in fact be predictable.

Despite these incursions, the first known attempt to describe this new
branch of knowledge was due to Quetelet who, in 1835, coined it
\textquotedblleft Social Physics\textquotedblright . The notion encountered
support in Auguste Compte (1798-1857), who has for the first time considered
it as a separate scientific discipline. This idea would be raised up again
by Majorana, in the 20th Century, in his seminal paper on the analogy
between statistical laws in Physics and in Social Sciences, where he
outlined the opportunities and drawbacks of applying methods of the former
to the latter [4]. Recently, some physicists prefer to call it
\textquotedblleft Phynance\textquotedblright\ in a contraction of the terms
\textquotedblleft Physics\textquotedblright\ and \textquotedblleft
Finance\textquotedblright\ [5,6], others have adopted the terminology
\textquotedblleft Financial Physics\textquotedblright ; however, the name
Econophysics has prevailed since it encompasses a much wider focus ranging
from Economics to Finance.

Although Econophysics has emerged from the urge of describing economics or
financial phenomena by means of applying methods from the science of
Physics, it is worthy to note that the first power-law ever discovered in
nature, a most commonly distribution evidenced in Physics,\footnote{%
Power-laws have received considerable attention in physics because they
indicate scale free behaviour and are characteristic of critical or
nonequilibrium phenomena.} was originally observed in Economics by Pareto
(for details, see Ref. [7]), when analyzing the distribution income among
the population. Similarly, Ref. [8] proposed the first theory of market
fluctuation, five years before Einstein's famous paper on Brownian motion
[9], where he derived the partial differential heat/diffusion equation
governing Brownian motion and made the estimate for the size of molecules.
Specifically, Ref. [8] gave the distribution function for the Wiener
stochastic process -- the stochastic process underlying Brownian motion --
linking it mathematically with the diffusion equation. It is thus telling
that the first theory of the Brownian motion was developed to model
financial asset prices! These two examples illustrate that the relation
between both sciences is bi-directional and not a one-way route, as one
might believe, a fact that must be considered when studying this subject.
Nevertheless, the impact of Physics in Economics/Finance is much more
evident than the reverse.

\section{Scope}

Before proceeding any further, additional enlightenment is required in order
to fully understand the meaning and implications of the word
\textquotedblleft Econophysics\textquotedblright . There are, indeed, a
number of attempts in literature to define it. Ref. [10, p. 3] considers
that \textquotedblleft Econophysics is a hybrid discipline (\ldots ) that
applies various models and concepts originated in Physics to economic (and
financial) phenomena.\textquotedblright\ and adds, in a more eloquent way,
that \textquotedblleft Econophysics presents itself as a new way of thinking
about the economic and financial systems through the `glasses' of
physics\textquotedblright . According to Ref. [11, p. 1] this discipline is
\textquotedblleft a quantitative approach using ideas, models, conceptual
and computational methods of Statistical Physics\textquotedblright . From
Mantegna and Stanley point of view [12, p. 355] \textquotedblleft the word
Econophysics describes the present attempts of a number of physicists to
model financial and economic systems using paradigms and tools borrowed from
Theoretical and Statistical Physics\textquotedblright .

Basically, the general idea is to use concepts and tools of Physics in order
to study economic/financial problems [10-12]. Building on this, physicists
have been mainly applying concepts and methodologies of Statistical Physics (%
\textit{e.g.}, scaling, universality, disordered frustrated systems and
self-organized systems) to describe such complex systems as economic or
financial systems, as they appear to be. Indeed, most approaches based on
the fundamentals of Physics perceive financial/economic phenomena as complex
evolving systems [13]. This is due to the multiple interacting components
exhibited by the inherent time series, \textit{e.g.}, stock market indexes
or inflation rates. In particular, these systems are expressed in the light
of their statistical properties and their principles (microscopic models,
scaling laws) are used to develop models to explain the corresponding
behaviour [10]. Some examples of the application of Statistical Physics to
Finance can be found in [14-18].

\section{The call for a new discipline}

One fundamental question that may arise when approaching this subject is
what has triggered the urge for this new discipline. It seems reasonable
that if Econophysics has emerged some underlying factors must have been
driving it. When addressing this matter two main reasons seem to apply: (i)
the limitations of the traditional approach of Economics/Finance and, (ii)
the advantages of the empirical method used in Physics. As for the first,
there was already a growing debate in literature about the drawbacks of
methods advocated by Economics/Finance. Recently, some of these shortcomings
were listed by Ref. [19] in the economic domain. According to the author
concepts like utility maximization, perfect competition and diminishing
marginal productivity are empirically and logically flawed and should not be
used in Econophysics. For instance, a foundation belief of neoclassical
Economics is that all individuals want to maximize their subjective utility.
However, a considerable number of violations exist that contradicts the
neoclassical theory of consumer. One relates to income, another has to do
with habit, and a third one refers to the way tastes are formed, transmitted
and modified.

Another assumption frequently criticized is the postulate of perfect
competition. Indeed, numerous researchers have found that marginal cost,
denoted by the marginal benefit of consumption, is irrelevant to the firm
[20]. Finally, the proposition of diminishing marginal productivity is also
falsified.

Alternatively, critics to the Finance theory refer basically to the
Efficient Market Hypothesis (EMH) formulated by Ref. [21], which comprises
three major versions: "weak", "semi-strong", and "strong" form. Weak EMH
claims that prices on traded assets (\textit{e.g.}, stocks, bonds, or
property) already reflect all past publicly available information.
Semi-strong EMH both asserts that prices reflect all publicly available
information and that prices instantly change to reflect new public
information. Strong EMH additionally postulates that prices instantly
reflect even hidden or "insider" information. Despite its popularity, this
principle is strongly controversial and has been successively questioned,
since it represents a mere idealization that can hardly be verified [22-24].
In fact, the idea that markets are rational, from which this theory departs,
is a theoretical construction that can be easily violated. Another example
stands from the Capital Asset Pricing Model (CAPM), which can not be applied
(i) if investors differ in their expectations and, (ii) if they cannot
borrow limitless amount of money at the same interest rate.

As opposed, the appeal from Physics relies on the kind of methodology
frequently applied, mainly focused on an experimental basis. Thus, while
economists often start with a model and after test what the data can say
about such model, Physics tries to unfold the empirical laws which one later
models. In other words, while the driving force in Physics is the quest for
universal laws, economists are much more concerned about documenting
differences. According to Ref. [25] the need for an alternative approach
stems from the fact that economic models cannot fully characterize the real
behaviour of stock markets returns. This is especially true as it is widely
recognized that the fundamental laws governing economic or financial systems
were not yet completely understood. In addition to this, we may also stress
that the empirical regularities exhibited by economic phenomena suggest that
an important part of the social order may be incorporated in the Physics
conceptual framework.

To sum up, we may say that the interest of physicists in economic/financial
arena is due to four main factors [26]: (i) economic fluctuations affect
everybody, which means that their implications are ubiquitous; (ii) everyone
would be affected by a breakdown of the World-wide financial system; (iii)
it is possible that methods and concepts developed in the study of
fluctuation systems might yield new results; and, (iv) the existence of
large data sets in economic/financial domain, which in some cases contains
hundreds of millions of events.

\section{The insight through Econophysics}

During its short existence many empirical research has been conducted
spanning different areas of knowledge in Economics and Finance. Space
limitations however restrict us to a brief overview of both of them,
centering the main focus on Finance Theory and in the interdisciplinary
application of the concept of entropy. Nonetheless, a short summary of the
state of the art is presented in order to provide a review of the kind of
research performed. Specifically, topics in Economics include the
distribution of income, the utility function, theories of how money emerges
and the application of symmetry and scaling to the functioning of markets
[27-32]. Among the important issues currently been debated in the World of
Finance, the stock prices fluctuations have been recurrently addressed.

Several reasons have been advanced for that [33]. Firstly, it may be
difficult to explain large fluctuations of asset prices based only on the
information about the fundamental economic factors. This may lead to a lack
of confidence on equity markets with the correspondingly consequences on
their liquidity. Secondly, volatility is an important factor in determining
the probability of bankruptcy of individual firms. Thirdly, price
fluctuations can strongly influence the bid-ask spread. Therefore, the
higher stock prices volatility, the higher that bid-ask spread will be; thus
affecting market's liquidity. Fourthly, hedging techniques may be affected
by the level of volatility, with the prices of insurance increasing with the
volatility level. Fifthly, an increase in risk associated with a growing
volatility may imply a reduced level of participation of investors in
economic activity with adverse consequences on investment. Last, an
increasing volatility may induce regulatory agencies to force firms to
allocate a larger amount of capital to cash equivalent investments, which
compromises the principle of capital efficient allocation. Additionally, the
actual Worldwide crisis, which started in 2008, with the subprime credit
default, has also drawn the attention to equity markets, their functioning,
the degree of stock prices oscillations and their implications in real
economy.

One major insight in Econophysics is the use of the Hurst exponents to
determine whether asset prices might exhibit long-range correlations and
thus may need to be described in terms of long-memory processes, such as the
fractional Brownian motion (FBM). The idea of using the FBM to model asset
price dynamics dates back to [34]. Since then, the Hurst exponent has been
calculated for many financial time series, such as stock prices, stock
indexes and currency exchange rates [35,36]. While studying market indexes
an interesting feature appears to emerge [36]: large and more developed
markets usually tend to be \textquotedblleft efficient\textquotedblright\
with a Hurst exponent close to $0.5$, whereas less developed markets tend to
exhibit long-range correlations. Accordingly, the multi-fractal dimension of
stock markets returns has also been addressed by Refs. [37-40]. In a recent
work, Ref. [40] concluded that the DAX data series shows very complex
self-similar structures, which may escape to any unique multi-fractal
description. One likely reason is that real financial data series are
super-positions of series with different properties. In the same line, Ref.
[41] found out that the multi-fractal structure of the traded volume of
equities encompassing the Dow Jones 30 arises essentially from the
non-Gaussian form of the probability density functions and from the
existence of nonlinear dependencies.

Furthermore, the correlations among stock returns have also been addressed
by means of the random matrix theory (developed in nuclear physics). It
seems that the problem of interpreting the correlations among large amounts
of spectroscopic data on the energy levels, whose exact nature is unknown,
is similar of interpreting the correlations among different stocks returns.
Therefore, with the minimal assumption of a random Hamiltonian, given by a
real symmetric matrix with independent random elements, a series of
predictions can be made. Some other examples concerning the application of
this methodology to Finance problems can be found in [42-44].

Another striking resemblance that unfolds when analyzing stock market
volatility is its resemblance with the turbulence in fluids. Ref. [12, p.
88] addresses this as follows: \textquotedblleft In turbulence, one ejects
energy at a large scale by, \textit{e.g.}, stirring a bucket of water, and
then one observes the manner in which the energy is transferred to
successively smaller scales. In financial systems `information' can be
injected into the system on a large scale and the reaction to this
information is transferred to smaller scales -- down to individual
investors\textquotedblright . Other studies dealing with turbulence include
Refs. [45-47]. Moreover, the Omori law for seismic activity after major
earthquakes has equally proved to be useful when understanding large crashes
in stock markets [48]. Examples may ever continue with the application of
some other concepts of Physics to financial markets, such as, the diffusion
anomalous systems, whose general framework can be provided by the nonlinear
Fokker-Planck equation [17], etc. There is, indeed, a great deal of other
empirical research using methods and analogies borrowed from Physics that
space limitations prevent us to describe any further.

However, one that has drawn considerable attention for its apparent ability
in describing stock market fluctuations is the concept of entropy. This
notion was originally introduced in 1865 by Clausius to explain the tendency
of temperature, pressure, density and chemical gradients to disappear over
time. Building on this, Clausius developed the Second Law of Thermodynamics
which postulates that the entropy of an isolated system tends to increase
continuously until it reaches its equilibrium state. Later, around 1900,
within the framework of Statistical Physics established by Boltzmann and
Gibbs [49,50], it was defined as a statistical concept. Around the middle of
the $20^{th}$ Century, it found its way in engineering and mathematics,
through the works of Shannon [51] in information theory and mathematics, and
of Kolmogorov [52] in probability theory. Significant research has been done
ever since with Shannon entropy providing the general framework for the
treatment of equilibrium systems where short/space/temporal interactions
dominate.

However, many systems exist that do not satisfy the simplifying assumptions
of ergodicity and independence. Some of these anomalies include [53]: (i)
meta-equilibrium states in large systems involving long range forces between
particles; (ii) meta-equilibrium states in small systems (100-200
particles); (iii) glassy systems; (iv) some classes of dissipative systems,
(v) mesoscorpic systems with non-markovian memory. Due to the prevalence of
these phenomena several entropy measures were derived. Among them, a most
popular one was Tsallis entropy [54], which constitutes itself as a
generalized form of Shannon entropy. Although first introduced by Havrda and
Charv\'{a}t [55] in cybernetics and late improved by Dar\'{o}czy [56], it
was Tsallis who has explored its properties and placed it in a physical
setting. Hence, it is also known as Havrda-Charv\'{a}t-Dar\'{o}czy-Tsallis
entropy.

Despite the debate generated over its meaning, for which the profusion of
several mathematical constructions has certainly played a central role,
entropy is commonly understood as a measure of disorder, uncertainty,
ignorance, dispersion, disorganization, or even, lack of information.
Recently, Ref. [57] has given it an econometric meaning, while considering
that the entropy of an economic system is a measure of the ignorance of the
researcher who knows only some moments' values representing the underlying
population. Besides its multiples applications, entropy has recently started
to be perceived as a consistent alternative to the standard-deviation, when
assessing stock market volatility.

The underlying rationality is that, as a more generalized measure, entropy
is able to capture uncertainty regardless of the kind of the empirical
distribution evidenced by the data. This is especially so, as it is widely
recognized that returns are usually non-normally distributed, where the
application of the standard-deviation turns out to be unsatisfactory. Ref.
[58] emphasizes that as a function of many moments of the probability
distribution function, entropy considers much more information than the
standard-deviation. Some of the main potentialities of this measure were
summarized by Refs. [59,60]: (i) it can be defined either for quantitative
or qualitative observations; (ii) whereas entropy depends on the potential
number of states of a distribution it is a result of the specific weight of
each state, and; (iii) the information value is related to the respectively
distribution function. For some empirical research concerning the
application of entropy in the financial domain the interested reader is
referred to [15,18].

At the end, let us stress that this constitutes a brief presentation of the
current efforts in Econophysics, mainly focused in the achievements in
Finance Theory, which has only illustrative purposes and cannot be
exhaustive.

\section{Opportunities \textit{versus} limitations}

More than a decade passed since Econophysics formally emerged as a new
discipline by the effort of H. E. Stanley and his colleagues, who are
frequently referred as the Boston group. During this period an intense
debate took place with a plethora of new papers and contributions arising
from both domains -- Physics and Economics/Finance -- as the above-mentioned
review illustrates. In order to assess the extent to whether this research
has spread, its main implications, lessons learnt, major insights and
whether the ultimate goals were achieved, an assessment is due. In doing so
an examination of opportunities and limitations is conducted. The aim is to
determine whether this approach is effective in describing complex
economic/financial phenomena for which the traditional approach has proven
to be insufficient.

By using the methodology applied in Physics which is extensively documented
in literature, Econophysics can go forward and facilitate an integrated
analysis based on the insights of both disciplines, with major benefits to
the global comprehension of economic/financial phenomena. This may be done
within the conceptual framework of Physics but without neglecting the
concepts, theories and paradigms which have been demonstrated to be
effective in the fields of Economics and Finance. With this kind of
combination Econophysics becomes much more powerful and gives a full
comprehension of the real implications of phenomena, which constitutes its
main challenge/opportunity. This is strictly in line with the universality
shown by the general use of Physics framework, which emphasizes that a
certain order may exist in nature that makes this generality possible. Thus,
there are reasons to believe that these approaches must be complementary and
not opposed, as many times it appears to be. This leads to a common
limitation frequently mentioned when addressing this matter, which derives
from the fact that if the traditional approach provided by Economics/Finance
is not found fully satisfactory, thus everything else in the field must be
disregarded. Although the mainstream work in Economics/Finance has
repeatedly been criticized, some progress has been made to understand how
Social and the World of Economics works, which must be considered when
addressing these issues. The consequence of denying such achievements is a
duplication of efforts that cannot be fruitful and at best can be a
fragmented view of reality.

A pitfall that can arise when seeing through Econophysics \textquotedblleft
glasses\textquotedblright\ is the tendency to recapitulate existing
theories, already developed in Physics, without adding some fresh
contributions. There may already appear some overlaps in previous works! The
general idea is that Econophysics should adapt the conceptual framework of
Physics instead of simply depicting it when processing economic/financial
data. This is a corollary of the above-mentioned principle according to
which Econophysics should combine both sciences. The extensive evidence of
power-laws in Economics/Finance may illustrate that tendency. In fact, the
ubiquitous nature of objects as fractal or self-similar has already been
criticized by Ref. [61], who found out that the \textquotedblleft scaling
range of experimentally declared fractality is extremely
limited\textquotedblright\ and by Ref. [62], who alerted that
\textquotedblleft one should be careful in not seeing a power-law decline in
each and every collection of data points with negative
slope\textquotedblright .

Furthermore, the impossibility in performing experiments due to the kind of
available data in Social Sciences constitutes an additional limitation that
can be partly suppressed by applying the Physics methodology based on the
quest for empirical laws. Even though experiments are still difficult to
undertake, by focusing on what the observations can reveal about the
phenomena and not trying to fit a particular model to the observed data,
Econophysics can effectively contribute to the understanding of Economics
and Finance World. A final shortcoming that should be pointed out is the
tendency of econophysicists to develop theoretical models essentially based
on the principles of Statistical Physics. If Physics can contribute to
Economics and Finance a question that may arise is why not try other
approaches based on different areas of Physics to see whether they can also
give an effective contribution?

By highlighting these limitations we have no intention to particularly
emphasize the potential flaws of Econophysics but rather to give a global
perspective on its main drawbacks in order to raise a debate that can only
be fruitful to the emergence of this new discipline. Accordingly, the
discussion must be seen in a constructive way since it may generate future
improvements in the field.

\section{Future directions}

Bearing on the above diagnostic some reflections arise. At the end, the
question initially addressed of whether this would be a fashionable topic
that turned out to be attractive for its innovative perspective but without
any substance seems to be even more fundamental. In other words, what is in
stake is if this is a new science or just a fashion trend that will
gradually disappear over time when the subject is no longer a novelty. In
view of the literature produced during these 15 years and on the advantage
of the experimental method defended by econophysicists, much more focused on
the data than on building a perfect elegant model, we might be tempted to
answer that a new science is emerging. Indeed, we believe that the
principles and the methodology, more than the particular techniques, might
be applied (\textit{e.g.}, Hurst exponents; random matrix theory), that can
really make the difference and provide new achievements. This idea is
corroborated by Ref. [10] through the neopositivist argument. In his view,
Econophysics can be considered a separate discipline and not merely a branch
of Economics since it proposes a different methodology based on a logical
empiricism and on the idea that observations are the core of all scientific
research. This is basically what the philosophical movement called
\textquotedblleft neopositivism\textquotedblright\ postulates: observational
evidence is indispensable for the knowledge of the World.

Let us conclude by mentioning that even though Econophysics is an emerging
discipline, it may aspire to be effective in describing Economics or
Financial systems if its principles and methods can mature. Lot of work is
going on in this field. The challenge is to see whether (or to what extent)
this is achieved in the coming years.

\end{document}